**Off-stoichiometry engineering of the electrical and optical properties of SrNbO$_3$ by oxide molecular beam epitaxy**

*Jasnamol Palakkal$^{1,2}$\*, Alexey Arzumanov$^2$, Ruiwen Xie$^2$, Niloofar Hadaeghi$^2$, Thomas Wagner$^3$, Tianshu Jiang$^2$, Yating Ruan$^2$, Gennady Cherkashinin$^2$, Leopoldo Molina-Luna$^2$, Hongbin Zhang$^2$, Lambert Alff$^2$*

$^1$*Institute of Materials Physics, Georg-August-University of Göttingen, 37077 Göttingen, Germany*

$^2$Institute of Materials Science, Technical University of Darmstadt, 64287 Darmstadt, Germany

$^3$*Quantum Design GmbH, 64319 Pfungstadt, Germany*

\*E-mail: jpalakkal@uni-goettingen.de

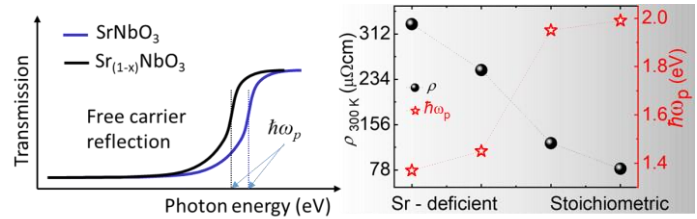

**TOC**

**Abstract**: The highly conducting and transparent inorganic perovskites Sr$B$O$_3$ with V, Nb, Mo, and their mixtures at the *B*-site have recently attracted the attention of the oxide electronics community as novel alternative transparent conducting oxides. For different applications from solar cells to transparent electronics, it is desirable to tune the optical transmission window in the ultraviolet (UV), visible and infrared (IR) range. The conventional approach is substitutional design at the *A*- and/or *B*-site. Here, we suggest a method by engineering the off-stoichiometry of the perovskite, opening new pathways to broaden the range of applications without adding additional elements. For oxide molecular beam epitaxy grown SrNbO$_3$ on GdScO$_3$ substrates, we show that controlled Sr deficiency shifts the plasma edge from about 2 eV in the visible range into the near-infrared region, 1.37 eV (similar to stoichiometric SrVO$_3$). Here, epitaxial growth allows going beyond the limitations of phase stability set by thermodynamics. The suggested approach opens a new design toolbox by including controlled vacancy sites as quasi-substitutional virtual elements.

**1. Introduction**

Electrically conducting and optically transparent materials, commonly known as transparent conducting oxides (TCO), are a scientifically important class of materials for basic materials research and are technologically significant for many optoelectronic devices. Single-crystalline and epitaxial thin films of 3$d$-, 4$d$-, and 5$d$- perovskites Sr$B$O$_3$ ($B$ = V, Nb, Mo, and Ta) are examples of TCOs. [1-5] A high charge carrier density offered by the unpaired $d$-electrons at the $B$-site cation makes them metallic, while the electron correlation and band transition shift their plasma frequency ($\omega_p$) towards the infrared region. [1-3] Traditionally known TCOs like Indium-Tin-Oxide (ITO) are wide bandgap semiconductors with high metallic conduction provided by effective carrier doping. [6] Since ITO is a wide-band gap material, the transparency resulted from the absence of any intraband transitions in the visible light spectrum. Design principles of TCOs without doping on transparent materials were suggested by changing the plasmonic frequency of metallic materials by altering their electron-electron correlation along with interband and intraband transitions. [1-3, 6] This way, these metals retain their metallicity while setting forth the optical transparency.

Both the conductivity ($\sigma$) and $\omega_p$ are proportional to the effective carrier concentration and inverse of the effective mass of the charge carrier. [7] In correlated metallic perovskites, the electron-electron correlation increases the effective mass, thereby decreasing the plasma frequency below the visible light spectrum (1.7-3.2 eV). [7] With the free carrier reflection edge, $\omega_p$, lying in the IR region and the interband optical absorption edge ($\omega_{Inter}$) in the UV region, a low-absorption region between $\omega_p$ and $\omega_{Inter}$ was reported in Sr$B$O$_3$. [1, 3, 8] A reduction of $B$-O-$B$ overlap by modulating the octahedral ($B$O$_6$) tilt was also reported to be shifting $\omega_p$ of Sr$B$O$_3$ by effectively replacing Sr by Ca.[2, 9] The effective mass of charge carriers is higher for Ca$B$O$_3$ compared to Sr$B$O$_3$. [9] Sr$B$O$_3$ was reported to possess a high IR transmittance and a low UV transmittance for $B$ = V, low IR transmittance and high UV transmittance for $B$ = Mo, and high IR-visible-UV transmittance for $B$ = Nb. [3] Enhanced correlation strength, as well as transition energy from O-2$p$ to $B$-$d$ orbitals, were reported to make the SrNbO$_3$ transparent over a wide spectrum. [3]

Resistivity ($\rho$) values in the metallic conducting range for SrNbO$_3$, vary from 28 to 970 µΩcm with a carrier density of ~2 x 10$^{22}$ cm$^{-2}$. [3, 10-15] A transition from metallic to insulating was observed in Sr$_{1-x}$$B$O$_{3-\delta}$ due to the presence of surface/deep defects appearing as a result of multivalent $B$-site cations. [4] An oxygen-reduced lattice offers fewer defects in these materials, stabilizing the metallic state which is useful for electrochemical devices. [4, 16] SrNbO$_3$ and SrNbO$_2$N were also found to behave as plasmonic photocatalysts and cause water splitting under sunlight. [17, 18] The plasmon excitations by phonons are responsible for the photocatalytic activity in SrNbO$_3$. [17] Wan *et al.* perceived that the photocatalytic activity of SrNbO$_3$ under visible to near-IR was due to hot electron generation from plasmon decay. [19] Tunable correlated plasmons were identified to exist in off-stoichiometric Mott-like insulating SrNbO$_3$ films. [20] The degenerate conduction band with a huge density of charge carriers of this large-bandgap material gives rise to different plasmonic effects. [19] Garcia-Castro *et al.* pointed out that nitrogen substitution can even make this perovskite a multiferroic. [21] Symmetry modifications by strain cause even a topological band structure in SrNbO$_3$. [11] Strained SrNbO$_3$ films in the extreme quantum limit behave as Dirac semimetals with ultrahigh mobility of Dirac electrons. [11]

It is no doubt, that SrNbO$_3$ is an interesting material for investigating fundamental materials properties, but it is also relevant for applications in optoelectronics. Vital is to keep the high conductivity of these materials while altering the optical transparency range. With this aim, we fabricated perovskite oxide SrNbO$_3$ thin films with different Sr:Nb ratios using the novel approach of off-stoichiometry engineering. Considering vacancies as quasi-substitutional elements, we will show that Sr vacancies have the same effect on the plasma edge as the complete replacement of Nb by Vanadium. We demonstrate the control over off-stoichiometry of the films under co-evaporation of elemental sources using oxide molecular beam epitaxy (MBE) technique[22] for tailoring the optical properties of Sr$_{(1-x)}$NbO$_3$ (SNO) thin films.

## 2. Results and Discussion

### 2.1. Controlled off-stoichiometric Sr$_{(1-x)}$NbO$_3$ (SNO) thin films

The Sr flux rate was carefully adjusted to deposit controlled off-stoichiometric SNO films on the GdScO$_3$ (110) (GSO) substrate buffered with a SrTiO$_3$ (STO) layer. SNO was deposited by oxide MBE, and STO was done by pulsed laser deposition as in our previous work. [23] Four SNO films with different Sr/Nb ratios were produced by varying the Sr flux rate from 0.12 Å/s to 0.09 Å/s while keeping all other parameters the same. Details of the deposition can be found in the experimental section and reference. [23]

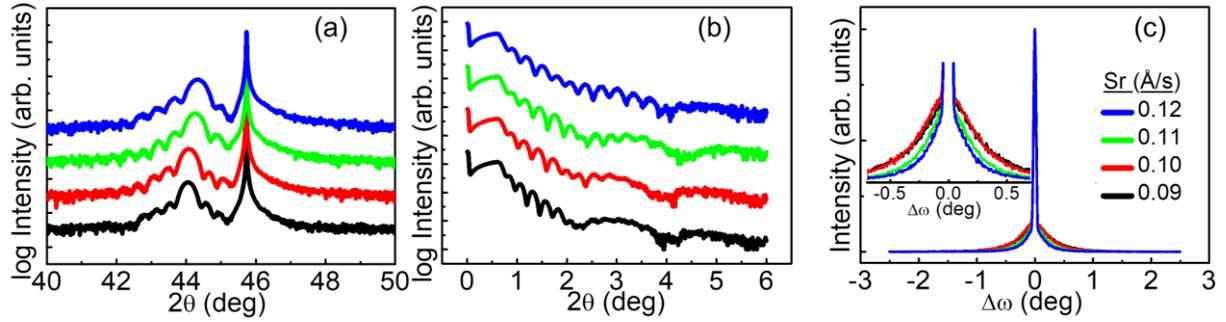

**Figure 1** (a) XRD pattern (b) XRR, and (c) rocking curve of SNO films deposited with different Sr-flux rates.

X-ray diffraction (XRD) pattern around GSO (220) reflection of the SNO films deposited with different Sr rates is shown in **Figure 1**a. The SNO (002) reflection with Laue oscillations is visible next to the substrate peak. This points to the coherent growth of the epitaxial SNO films on the orthorhombic GSO. The XRD pattern for a wide $2\theta$ range shows reflections corresponding to other parallel planes (see *supplementary* **Figure S1**). The position of the SNO (002) reflection is shifting towards the negative $2\theta$ direction as the Sr rate was decreased, showing an increase in the lattice constant with vacancy/defect formation.[24] The calculated out-of-plane lattice constant (*c*) from the XRD pattern is plotted as a function of the Sr flux rate in **Figure 3**c.

The X-ray reflectivity (XRR) image of the multilayered GSO/STO/SNO thin films shows Kiessig fringes, as can be seen in **Figure 1**b. This substantiates a sharp interface and homogeneous film growth with a low surface roughness of the films. The more significant periods in the XRR oscillations come from the STO buffer layer, the thickness of which is estimated to be 4 - 5 nm for all the films. The importance of using an STO buffer layer for improving the quality of the thin film growth was discussed elsewhere.[23] The smaller periods in the XRR are imminent from the SNO layer, whose thickness is estimated to lie between 20 and 25 nm for all the films. The same values are obtained from a thickness calculation using the Laue oscillations in the XRD, confirming the coherent crystal growth of the films. **Figure 1**c depicts the rocking curve of the films. As the flux rate of Sr is decreased from 0.12 Å/s to 0.09 Å/s, the coherency shows a decrease, associated with the formation of defects originating from off-stoichiometry.

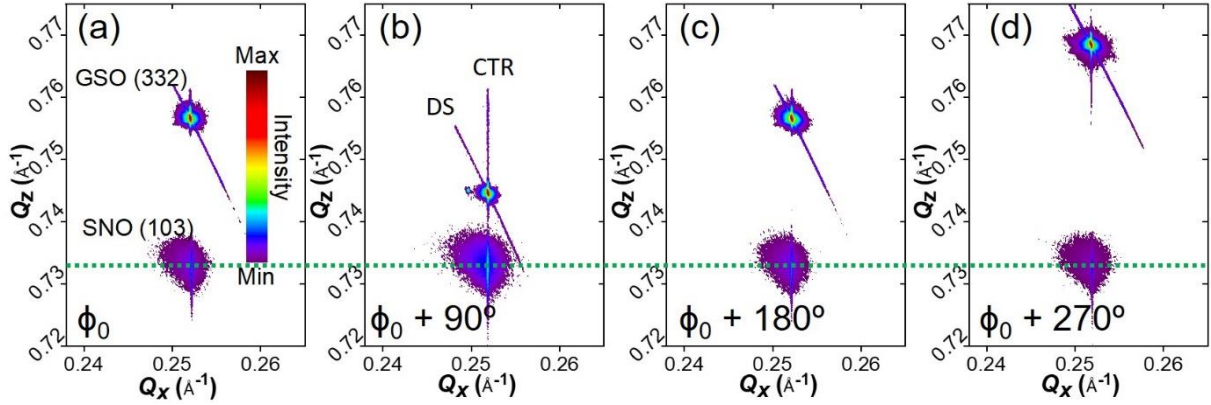

**Figure 2** The RSM images of the SNO films with Sr-flux rate of 0.12 Å/s with different azimuth (ϕ) angles (a) $\phi_0$, (b) $\phi_0 + 90º$, (c) $\phi_0 + 180º$, (d) $\phi_0 + 270º$. CTR and DS in (b) denote the crystal truncation rod and detector streak, respectively.

**Figure 2** shows the reciprocal space maps (RSM) of a SNO thin film with Sr = 0.12 Å/s measured around the GSO (332) and SNO (103) reflections at four azimuthal (ϕ) angles of 90º increment. In all four azimuthal angles, the *c* lattice parameter is the same, as denoted by the green dashed line, asserting the perpendicular SNO crystal growth on the orthorhombic GSO substrate. [23] The *c* is found to be 4.09 Å, as found from the XRD 2θ general scan. The film is in-plane locked to the substrate, with an in-plane lattice constant *a* of 3.97 Å, the same as that of the GSO substrate. The RSM images of other films are shown in **Figure S2-S4**. All the films are cubic, with the same *c* in all four angles of ϕ. The calculated *c* from RSM for these samples is the same as shown in **Figure 3**c. Upon decreasing the Sr flux rate, the films show a relaxation, as observed by the shift in the RSM contour to the negative $Q_X$ direction, displaying an increase in the *a*. The crystal truncation rod (CTR) and detector streak (DS) are noted in **Figure 2**b, which can be seen in all the RSM images. The CTR of the substrate and film lies at the same $Q_X$ value, suggesting that the Bragg angle in this direction (and thus *a*) for all the films is the same as that of the substrate. However, part of the film is relaxed, portraying defects in the system.

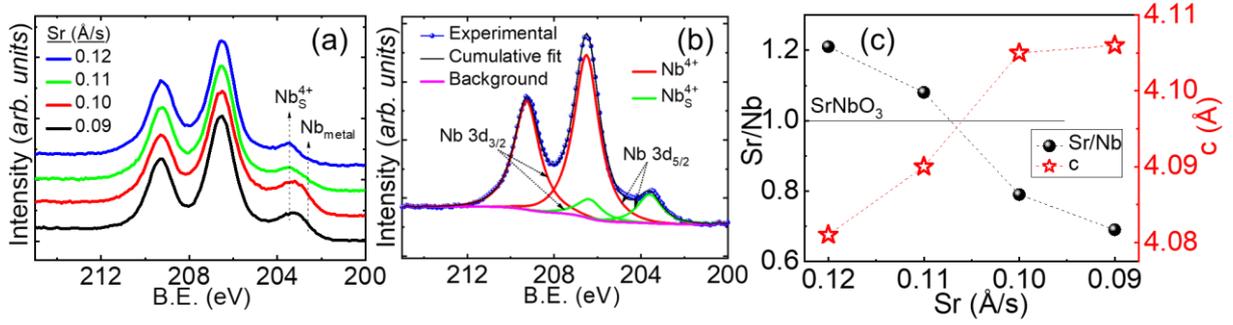

**Figure 3** (a) Nb 3*d* photoelectron spectra of SNO films. (b) The fitted Nb 3*d* photoelectron spectra of SNO film for Sr = 0.12 Å/s. (c) The Sr/Nb ratio and the out-of-plane lattice constant, *c*, plotted as a function of the Sr-flux rate.

**Figures 3**a and **3**b show the Nb 3*d* photoelectron spectra measured for SNO films and the fitted spectra for Sr = 0.12 Å/s, respectively. All films contain peaks conforming to the $Nb^{4+}$ oxidation state and the corresponding final state, the $Nb_s^{4+}$ spectral line, as discussed in our previous work.[25] Films with Sr = 0.10 and 0.09 Å/s show traces of metallic Nb ($Nb_{metal}$). The Sr/Nb ratio estimated from these XPS measurements is shown in **Figure 3**c as a function of the Sr rate. As expected, the Sr/Nb decreases with the decrease in the Sr flux. The quantitative stoichiometry determination using XPS in thin films (error 15-20 %) is not enough to unambiguously determine the charge compensation mechanisms due to the off-stoichiometry. Sr excess can be compensated by niobium deficiency and additional oxygen deficiency depending on how many Sr cations occupies a Nb site (a $Sr_{Nb}''$ defect in Kröger-Vink notation). Vice versa, Sr deficiency can be either compensated by $Nb_{Sr}^{\bullet\bullet}$ or $V_O^{\bullet\bullet}$ defects. The required amount of defect sites is at extreme Sr:Nb ratios in the range of 10%.

To check the microstructure, high-angle annular dark-field scanning transmission electron microscopy (HAADF STEM) were performed on Sr = 0.12 and 0.09 Å/s samples as depicted in **Figures S5** and **S6**. The data confirms high crystallinity thin films. Looking with high resolution into the atomic scale defect properties, we observe small regions with excess Sr in the stoichiometric (Sr = 0.12 Å/s) sample in an otherwise perfectly stoichiometric matrix (**Figures S5**a and **S5**b). For high Sr deficiency we find occasionally Nb double layers in Sr = 0.09 Å/s sample (see line profiles in **Figure S6**), i.e. an extended stacking fault. The density of these defects is too low to accommodate all Sr vacancies. Therefore, we conclude that the surrounding matrix is still highly Sr deficiency. Note that the observed stacking faults may contribute to reduce the resistivity, but are not relevant for the interpretation of the optical properties which are clearly due to the Sr vacancies in the $Sr_xNb_yO_3$ matrix, as also confirmed by the density functional theory (DFT) modelling described later.

## 2.2. Electrical and optical properties

### 2.2.1. DC Electrical transport properties

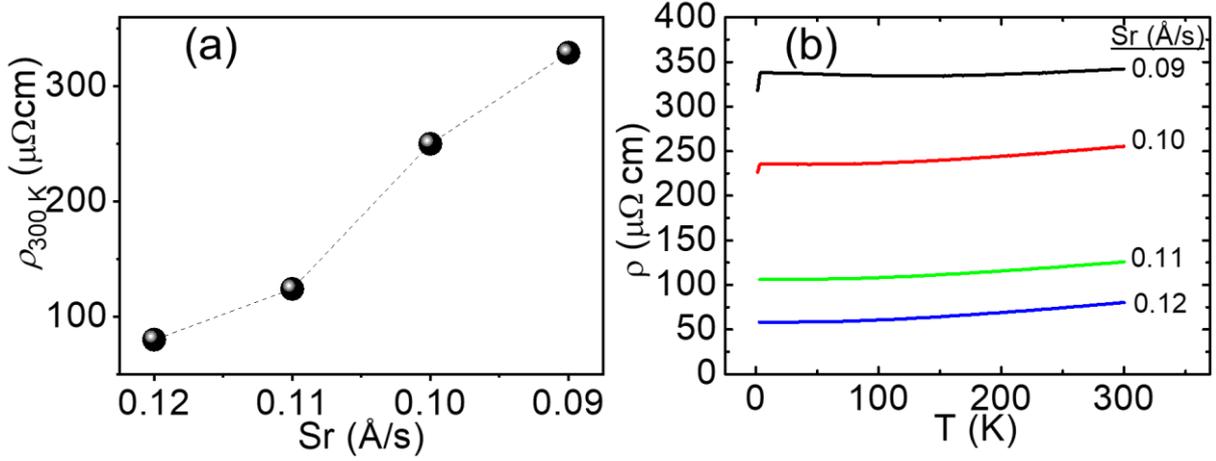

**Figure 4** Resistivity ($\rho$) measured (a) at room temperature and (b) as a function of temperature for the SNO films.

The resistivity ($\rho$) of the films at room temperature (**Figure 4**a) increases with a decrease in Sr rate due to off-stoichiometric defects introduced by Sr deficiency. The stoichiometric SrNbO$_3$ (Sr = 012 Å/s sample) shows a $\rho$ of 80 μΩcm at room temperature, and it reaches 329 μΩcm for Sr = 0.09 Å/s sample. The $\rho$ of all our SNO samples lies in the range of the $\rho$ values reported previously for SrNbO$_3$ ($\rho$ of 28 to 970 μΩcm).[3, 10-15]. An increased defect concentration induced by Sr deficiency decreases the mean free path, causing a reduction in the conductivity. The $\rho$ as a function of temperature is shown in **Figure 4**b. The $\rho$ follows typical metallic conduction behavior as a function of temperature. For the samples Sr = 0.10 and 0.09 Å/s, a minor downturn at low temperatures can be seen, which corresponds to the superconducting transition of the Nb metal, the presence of which was already identified in XPS.

### 2.2.2. Optical properties from spectroscopic ellipsometry

Spectroscopic ellipsometry was performed on the SNO thin films for UV, visible, and near-infrared (NIR) wavelengths. The extracted optical constants using Kramers–Kronig consistent B-spline model [26] are shown in **Figure 5**. **Figure 5**a shows the complex refractive index ($n$) and extinction coefficient ($k$). The real ($\varepsilon'$) and imaginary ($\varepsilon''$) parts of the complex dielectric constant ($\varepsilon$) are given in **Figure 6**b. The $n, k, \varepsilon'$, and $\varepsilon''$ decrease upon increasing photon energy and increase at higher energy ranges with optical transitions, following a typical Drude model for metals. For all the films, the $k$ appears to show a Drude peak below ~2 eV and the first interband transition above ~4 eV, as in the case of previous work on SrNbO$_3$. [19, 20] Conventional (bulk) and correlated plasmons were reported in Sr$_{1-x}$NbO$_{3+\delta}$. [20] It was reported that the correlated plasmons vanish, and only conventional plasmons exist when the Sr$_{1-x}$NbO$_{3+\delta}$ system is metallic. [20] A conventional plasmon was reported to occur at ~1.9 eV due to the increased free-charge carriers in metallic SrNbO$_3$ and correlated plasmons occurring at multiple plasmon frequencies viz. ~1.7, ~3.0, and ~4.0 eV in insulator-like Sr$_{1-x}$NbO$_{3+\delta}$. [20] The $\varepsilon'$ for a wavelength of 1000-245 nm are shown in **Figure S7**a. The loss spectrum plotted in terms of the loss function (LF) for a wavelength of 1000-245 nm is shown in **Figure S7**b. LF is calculated using the inverse of $\varepsilon$, as given by the equation LF = -Im[$\varepsilon(\omega)$], which is $\varepsilon''/(\varepsilon'^2 + \varepsilon''^2)$.[19] The Sr = 0.12 and 0.11 Å/s samples show peaks in the $\varepsilon'$, $\varepsilon''$, and LF around 3-4 eV, indicating the occurrence of correlated plasmons and a coupling between optical and plasmonic excitations, as reported before. [20] Such an absorption peak was also associated with the t$_{2g}$ to e$_g$ interband transition due to the hybridization of 2$p$ and 4$d$ states. [14, 27] This peak is absent in Sr-deficient samples, possibly due to the increased defect concentrations, as reported in SrNbO$_3$ films with sizeable defects. [14]

The wavelength corresponding to the plasma frequency ($\omega_p$), defined as the wavelength at $\varepsilon' = 0$, is 622 nm for the stoichiometric SNO (Sr = 0.12 Å/s), with a corresponding plasma energy ($\hbar\omega_p$) of 1.99 eV. With Sr-deficiency, the plasma wavelength increased to 634 nm ($\hbar\omega_p$ = 1.95 eV) for Sr = 0.11 Å/s, 852 nm ($\hbar\omega_p$ = 1.45 eV) for Sr = 0.10 Å/s, and 905 nm ($\hbar\omega_p$ = 1.37 eV) for Sr = 0.09 Å/s. Park *et al.* reported an $\hbar\omega_p$ of 1.98 eV and 1.82 eV for SrNbO$_3$ experimentally and theoretically, respectively. [14] The $\hbar\omega_p$ values for the Sr = 0.12 and 0.11 Å/s fall under that reported for metallic SrNbO$_3$ and decrease when the conductivity decreases in Sr-deficient samples, corroborating the rise of plasmons at these photon energies. [14, 19, 20]

The LF maximum (LF$_{max}$) decreases upon reducing the Sr content. The plasma energy found as energy at LF$_{max}$ ($\hbar\omega_{pLF}$) shows a slight blue shift for the stoichiometric samples and a considerable blue shift for the Sr-deficient samples. $\hbar\omega_{pLF}$ is 2.037 eV (for both Sr = 0.12 Å/s and Sr = 0.11 Å/s) and 1.93 eV (for both Sr = 0.10 Å/s and Sr = 0.09 Å/s). This blueshift indicates the scattering of free carriers and plasmon dephasing in the materials system. Here, the $\hbar\omega_p$ and $\hbar\omega_{pLF}$ can be associated with correlated and conventional plasmon energies, respectively. With electron-electron correlations, the effective mass of the charge carriers increases, causing an increase in the plasmon dephasing time and decreasing the plasma frequency. [7] The corresponding scattering time ($\tau$) and plasmon dephasing time ($T_2$) can be calculated from the blue shift using the equations $\hbar/\tau = \sqrt{(\hbar\omega_{pLF}^2 - \hbar\omega_p^2)}$ and $T_2 = 2\hbar/\tau$. The $\hbar\omega_p$, $\hbar\omega_{pLF}$, $\tau$, and $T_2$ are listed in **Table**1.

$\tau$ calculated using the equation $\tau = \hbar/\Gamma$ [10] ($\Gamma$ is the full-width half maximum of the curve with $LF_{max}$ as the peak) is 0.455 fs for the Sr = 0.12 Å/s sample, which is comparable with that calculated from the blue shift. Lorentz model was employed for obtaining $\Gamma$ from the fit. [10] The peak of LF spectra of other samples could not provide a good fit.

**Table 1** The plasma energy corresponding to correlated ($\hbar\omega_p$) and conventional ($\hbar\omega_{pLF}$) plasmons together with the scattering time ($\tau$) and plasmon dephasing ($T_2$) time calculated from the blue shift.

| Sample | $\hbar\omega_p$ (eV) (Correlated) | $\hbar\omega_{pLF}$ (eV) (Conventional) | $\tau$ from blue shift (fs) | $T_2$ from blue shift (eV) |
|---|---|---|---|---|
| Sr = 0.12 Å/s | 1.99 | 2.037 | 1.51 | 0.87 |
| Sr = 0.11 Å/s | 1.95 | 2.037 | 1.11 | 1.18 |
| Sr = 0.10 Å/s | 1.45 | 1.93 | 0.51 | 2.58 |
| Sr = 0.09 Å/s | 1.37 | 1.93 | 0.48 | 2.74 |

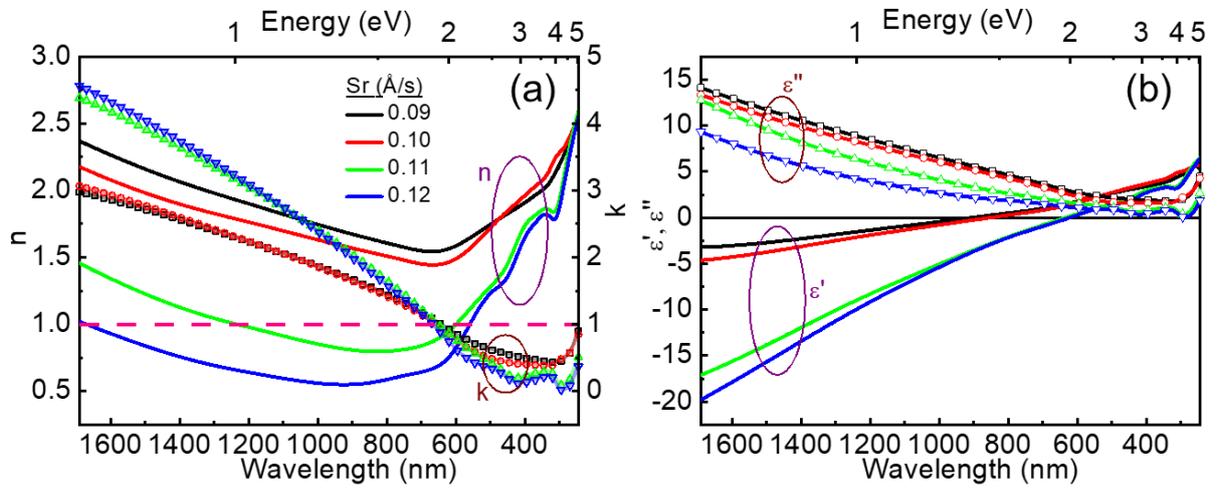

**Figure 5** (a) The complex refractive index ($n$) and extinction coefficient ($k$) and (b) the real ($\varepsilon'$) and imaginary ($\varepsilon''$) parts of the complex dielectric constant ($\varepsilon$), extracted for the SNO films from spectroscopic ellipsometry.

## 2.3. DFT modeling of the optical spectrum of stoichiometric and Sr-deficient SrNbO$_3$

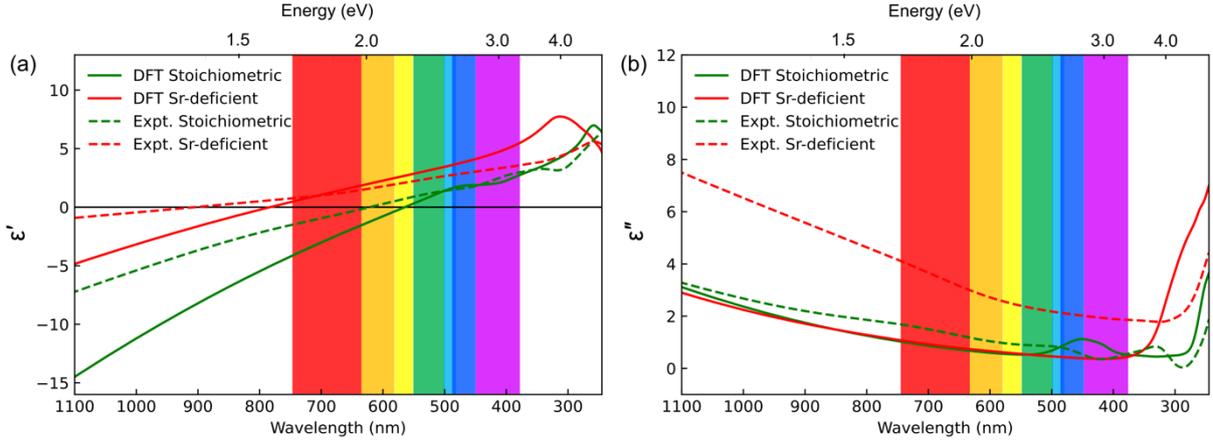

**Figure 6** (a) The real ($\varepsilon'$) and (b) imaginary ($\varepsilon''$) parts of calculated dielectric function ($\varepsilon$) by DFT (solid lines) corresponding to stoichiometric (green) and Sr-deficient (red) systems. The experimental measurements for stoichiometric (sample Sr = 0.12 Å/s) (green dashed line) and Sr-deficient (sample Sr = 0.09 Å/s) (red dashed line) are shown for comparison.

Motivated by the different optical properties experimentally observed in stoichiometric and Sr-deficient samples, we further performed DFT calculations of the dielectric function ($\varepsilon$) on both structures. **Figure 6**a shows the real part of the dielectric function $\varepsilon'$ as a function of wavelength. It can be noticed that the obtained plasmon frequencies are slightly overestimated, being $\hbar\omega_\mathrm{p}$ approximately 2.2 (1.99) and 1.6 (1.37) eV of DFT (experiment) results corresponding to the stoichiometric and Sr-deficient systems, respectively. Such overestimation can be understood if one takes electron-electron correlation effects into consideration.[14] That is, scaled by the inverse of the renormalization factor $Z_k$, the effective mass is increased with respect to the band effect mass. As a consequence, the plasma energy $\hbar\omega_\mathrm{p} = \hbar \sqrt{(e^2/\varepsilon_0)}\sqrt{(Z_k n/m_b^*)}$, is rescaled by a factor of $\sqrt{Z_k}$. For SrNbO$_3$, $Z_k$ is taken as ~0.72.[14] Accordingly, the values of 2.2 and 1.6 eV obtained by DFT calculations are reduced to 1.9 and 1.4 eV for stoichiometric and Sr-deficient systems, respectively, which fit well with experiments.

**Figure 6**b shows the imaginary part ($\varepsilon''$) of the complex dielectric functions, from which the optical absorption edge can be located. The position of the main absorption peak of SrNbO$_3$ given by current DFT calculation is around 5.9 eV, which agrees well with both the experiment and DFT results reported by Park *et al.*[14] In addition, an absorption peak at 2.8 eV is found from DFT results, while in the experiment the small absorption peak is observed around 3 to 4

eV. Such absorption feature is usually ascribed to the $t_{2g}$ to $e_g$ interband transition realized via the hybridization between $2p$ and $4d$ states. [27]

To elucidate the interband optical transitions, we show in **Figure 7** the calculated band structures, as well as the partial density of states (PDOS) of the stoichiometric and Sr-deficient crystal structures. We can see from **Figure 7** that by introducing Sr vacancies, the valence band maximum (VBM) consisting of mainly O $2p$ orbitals is increased to around -2.6 eV from -3.5 eV in the stoichiometric structure. Moreover, the conduction band minimum (CBM) of the Sr-deficient system is slightly enhanced (~ -0.7 eV) as compared to that of the stoichiometric case (~ -1.3 eV). The main differences displayed on the band structure explain the phenomenon that the main absorption peak in the Sr-deficient system initializes at relatively lower energy and possesses stronger intensity compared with the stoichiometric system (see **Figure** 6b). Note also that the experimentally measured $\varepsilon''$ show higher magnitudes than DFT results at the low-frequency range. We suspect that such discrepancy may be attributed to the simplification in modeling the Sr-deficient crystal structure. Other optical transition paths in a defective structure might exist that are hard to capture fully using a simplified simulation model. Nevertheless, the DFT calculations are able to capture the difference in the main absorption edge, as well as the change of plasma frequency, brought up by Sr deficiencies.

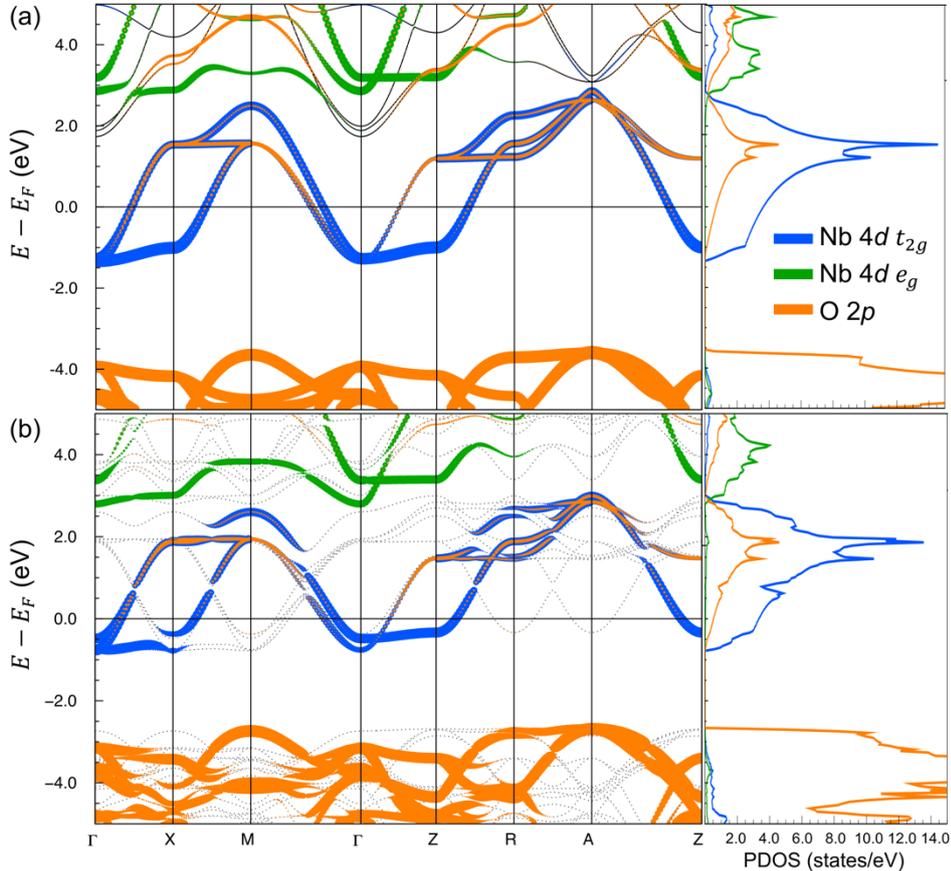

**Figure** 7. The band structure and partial density of states (PDOS) corresponding to (a) stoichiometric and (b) Sr-deficient models, respectively. Regarding the PDOS, the Nb 4$d$ $t_{2g}$ (blue), Nb 4$d$ $e_g$ (green) and O 2$p$ (orange) components are marked out.

## 3. Conclusions

We prepared off-stoichiometric $Sr_{(1-x)}NbO_3$ (SNO) films with different Sr: Nb ratios using oxide MBE. The lattice constant of the SNO shows an increase upon decreasing the Sr content due to vacancy formation. All the films have an $Nb^{4+}$ oxidation state with a $4d^1$ electronic configuration contributing an electron density of the order of $10^{22}$ cm$^{-3}$. The conductivity of the films improved when they became stoichiometric since the mobility of the charge carriers is enhanced in the stoichiometric film. The optical properties are also changing according to the stoichiometry. The plasma energy of 1.99 eV observed for the stoichiometric $SrNbO_3$ is decreasing with Sr-deficiency and reaching 1.37 eV for the Sr = 0.09 Å/s sample. Due to the high crystallinity in the stoichiometric sample, the interband transition from $t_{2g}$ to $e_g$ states was observed as an absorption peak in the complex dielectric permittivity around 2.7 eV photon energy. In this work, we achieved careful stoichiometry control using oxide MBE, thereby altering its plasma frequency from the visible light spectrum to the near-infrared region and making it transparent for all visible light without considerably changing its metallic conductivity. We conclude from a more general viewpoint that off-stoichiometry engineering is a useful tool to tailor materials properties in perovskites adding an additional toolbox for materials design for specific application requirements.

## 4. Experimental Section/Methods

*Thin film deposition:* The thin films were deposited using oxide MBE as explained in our previous work. [23] Compared to the previous report, [23] the Sr and Nb flux rates were different; moreover, we did not use any capping layer in the present work. Here, an Nb flux rate of 0.05 Å/s was used for the films and the four samples in the series were prepared by varying the flux rate of Sr from 0.12 to 0.09 Å/s. All other deposition conditions were the same as in the previous work. [23]

*Structural characterization:* The X-ray diffraction (XRD) measurements were performed using Rigaku Smartlab® diffractometer in a parallel beam geometry using a Ge(220)×2 monochromator (Cu Kα). A PHI Versaprobe 5000 spectrometer (Al Kα) was used for the XPS measurements.

X-ray photoelectron spectroscopy (XPS) measurements were performed in the Darmstadt Integrated System for Battery Research (DAISY-BAT). The XPS analysis chamber (base pressure <$10^{-9}$ mbar) is equipped with a PHI 5000 VersaProbe spectrometer (Physical Electronics) and a monochromatic Al Kα source ($hv$ =1486.7 eV). The diameter of the X-ray spot on the sample (illuminated area) was 200 μm. Photoelectrons were collected with the pass energy of $E_{pass}$ =23.5 eV at electron escape angle of 45° with respect to the surface normal. The background was subtracted by a Shirley-type function. The XPS results were used for a quantitative analysis of the elemental composition. For this purpose, the photoelectron peak positions and areas were obtained by a weighted least-square fitting of model curves to the experimental data. Note that a typical error in the XPS quantitative analysis is 15-20 %. The elemental composition was estimated considering a semi-infinite sample and assuming a homogeneous distribution of the elements over the depth.

High-angle annular dark-field scanning transmission electron microscopy (HAADF STEM) images were obtained using a Cs-corrected JEOL ARM 200F operating at a voltage of 200 kV. The convergence semi-angle was 29 mrad.

*Electrical and optical characterization:* The electrical properties on the thin film samples were performed with a van der Pauw geometry by means of sputtered gold contacts. A nanovoltmeter and a current source from Keithley Instruments were used together with a cryostat by Oxford Instruments NanoScience. A dual-rotating compensator ellipsometer RC2 from Woollam Co was used to perform spectroscopic ellipsometry for the UV-visible-near-infrared wavelengths.

*Computational methods:* The full-potential augmented plane waves and local orbitals' (FP-(L)APW+lo) method [28, 29], as implanted in the WIEN2k package [30] was utilized for the calculations of optical properties. In order to simulate the Sr-deficient structure, we constructed a 2 × 2 × 1 supercell and removed one Sr atom, thus obtaining $SrNb_{1.33}O_4$. The eexchange correlation of GGA-PBE [31] was selected. A 27 × 27 × 53 kmesh was used to converge the dielectric functions. The orbital-resolved DOS and band structure unfolding were performed using the full-potential local-orbital code FPLO [32] in which a kmesh of 30 × 30 × 30 was used.

**Supporting Information**

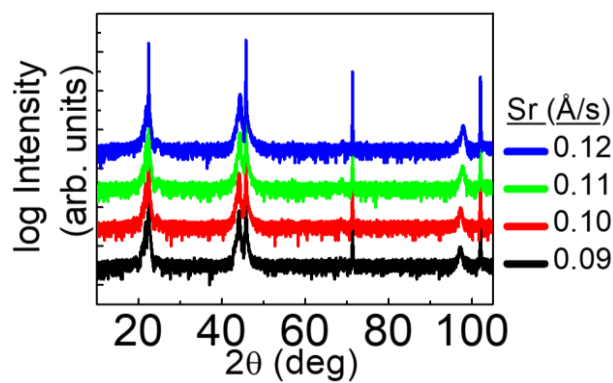

**Figure S1** XRD pattern of SNO films for a 2θ range of 10 - 105 deg.

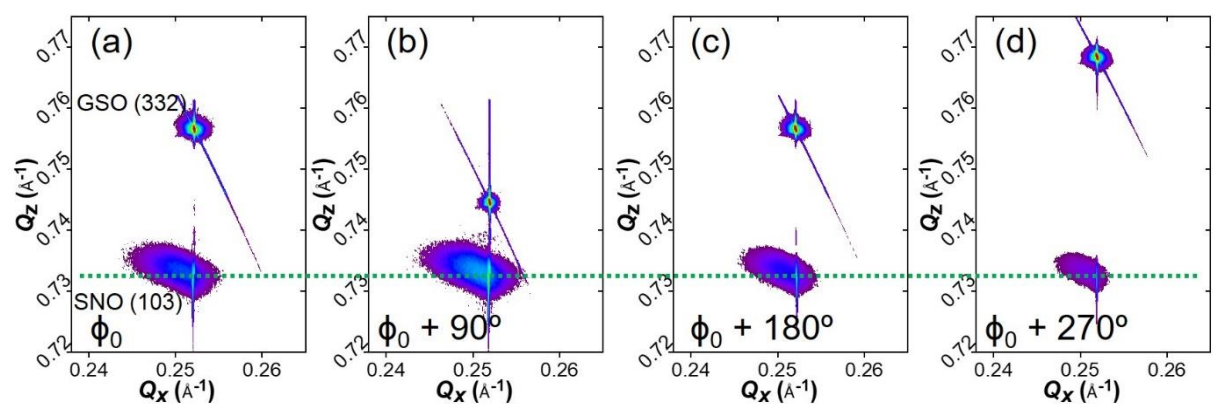

**Figure S2** The RSM images of the SNO films with Sr-flux rate of 0.11 Å/s with different azimuth (ϕ) angles (a) $\phi_0$, (b) $\phi_0 + 90°$, (c) $\phi_0 + 180°$, (d) $\phi_0 + 270°$

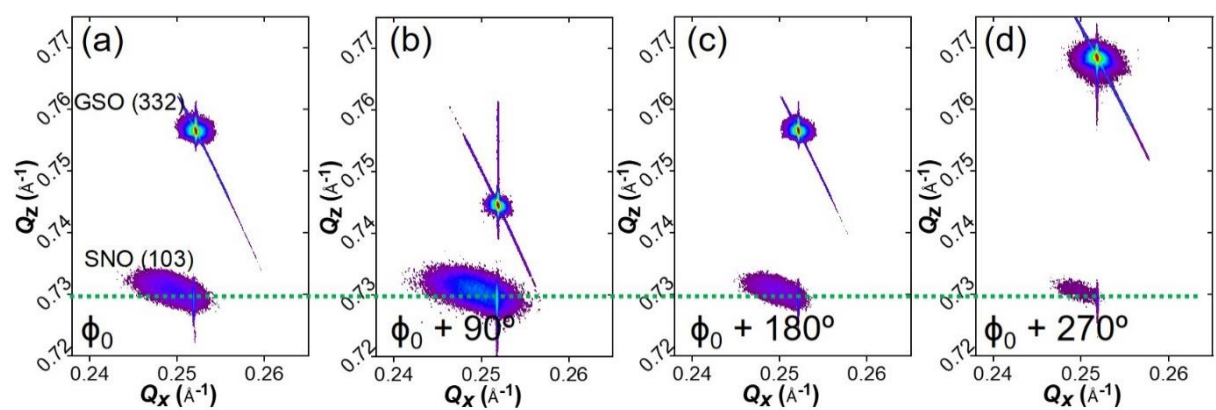

**Figure S3** The RSM images of the SNO films with Sr-flux rate of 0.10 Å/s with different azimuth (ϕ) angles (a) $\phi_0$, (b) $\phi_0 + 90°$, (c) $\phi_0 + 180°$, (d) $\phi_0 + 270°$

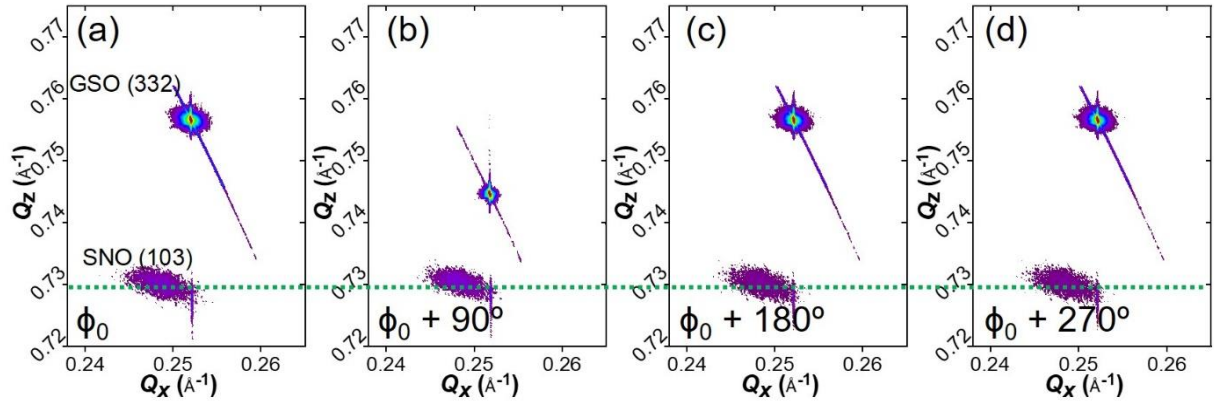

**Figure S4** The RSM images of the SNO films with Sr-flux rate of 0.09 Å/s with different azimuth (ϕ) angles (a) ϕ$_0$, (b) ϕ$_0$ + 90º, (c) ϕ$_0$ + 180º, (d) ϕ$_0$ + 270º

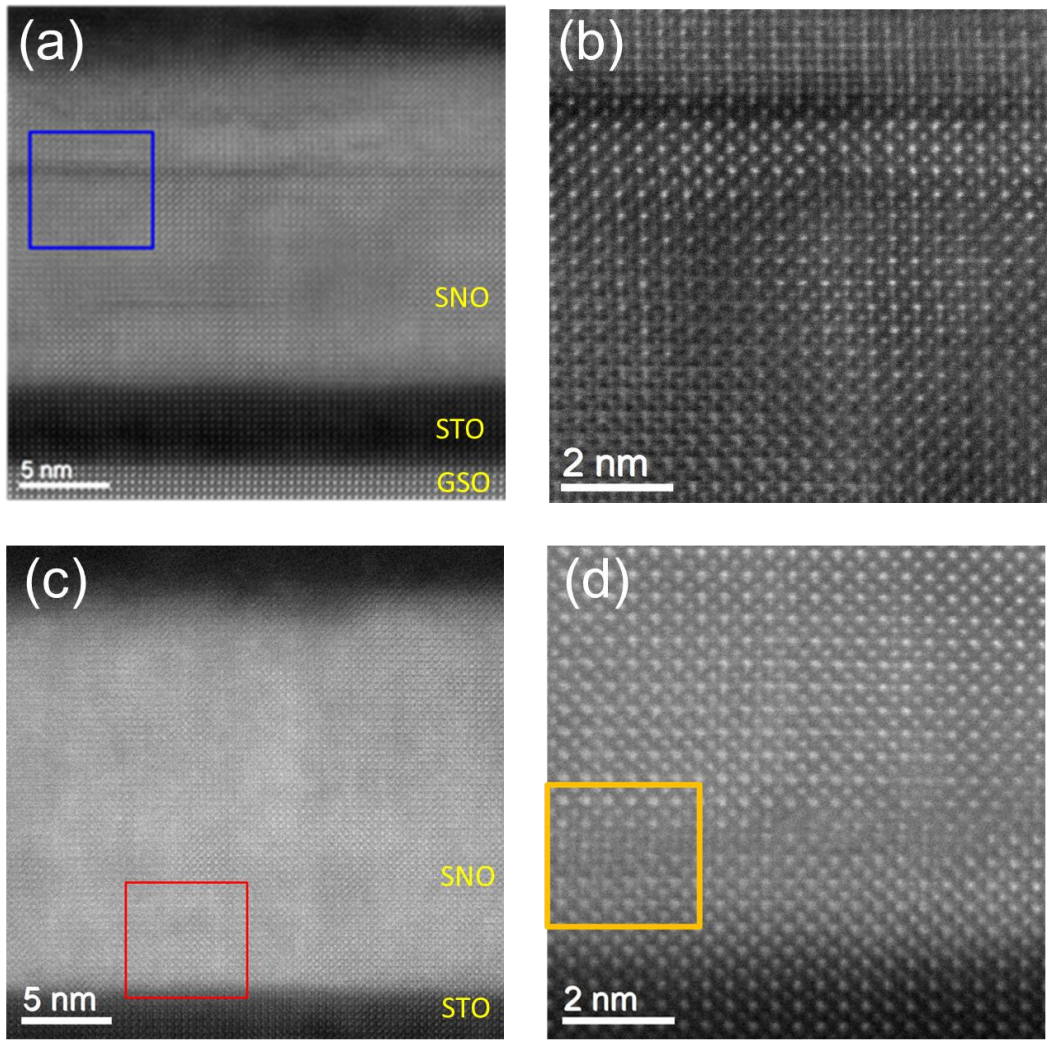

**Figure S5** (a) HAADF STEM image of the Sr = 0.12 Å/s and (c) Sr = 0.09 Å/s samples. (b) and (d) are the enlarged view of the area marked by the blue square in (a) and red square in (c) respectively.

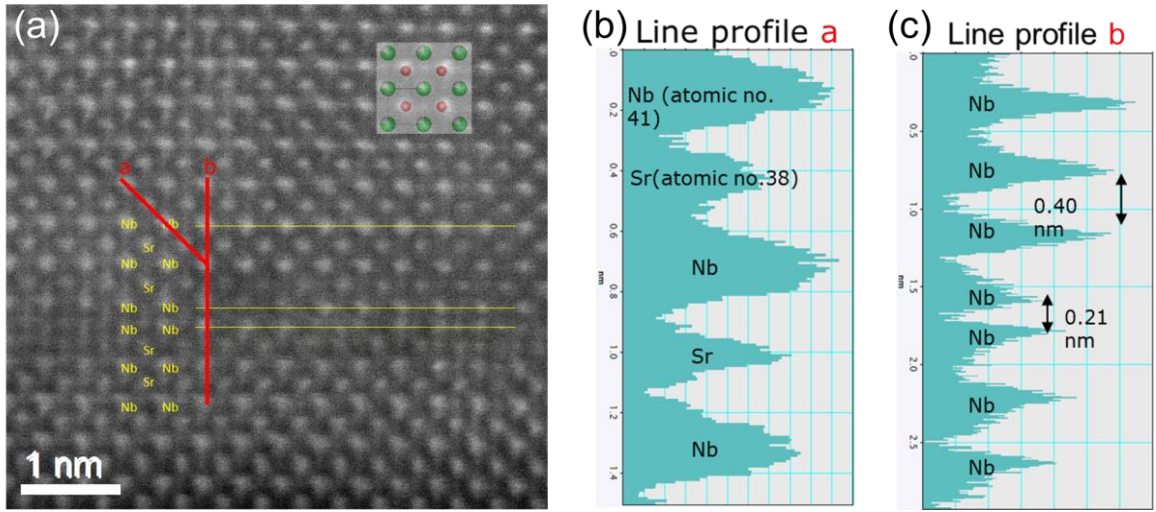

**Figure S6** (a) HAADF STEM image of the Sr = 0.09 Å/s sample, an enlarged view of the yellow square marked in Figure 4d. (b) and (c) are the line profiles corresponding to the a and b red lines marked in (a), respectively.

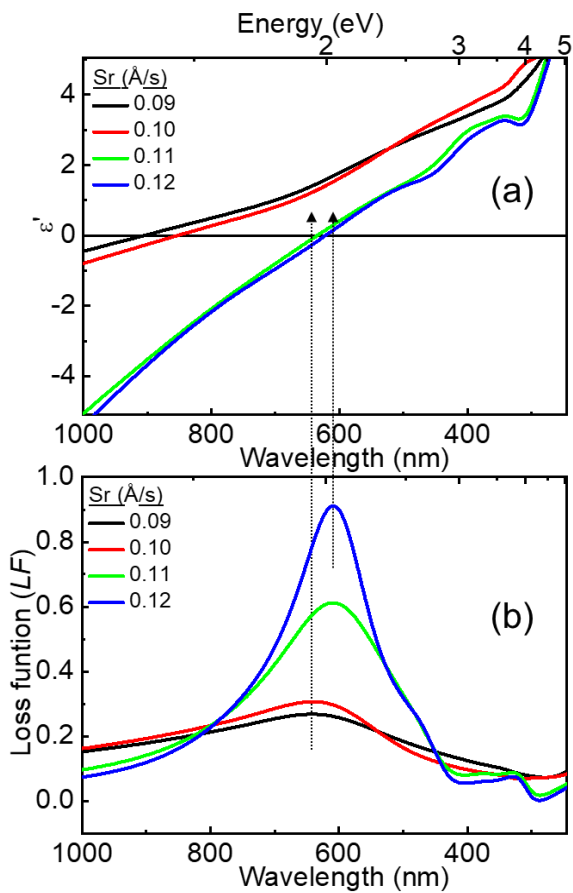

**Figure S7** (a) The real ($\varepsilon'$) part of the complex dielectric constant ($\varepsilon$) and the loss function (LF) of the SNO films for a wavelength range of 1000-245 nm.

**Acknowledgements**

This work was supported by Deutsche Forschungsgemeinschaft (DFG, German Research Foundation) under project 429646908. J.P. acknowledge the funding received from DFG-SFB 1073. T.J. and L.M.-L. acknowledge the European Research Council (ERC) under Grant Nos. 805359-FOXON, 957521-STARE and 101088712-ELECTRON. We thank Gabriele Haindl (thin film) and Laurin Elias Bierschenk (RSM), TU Darmstadt for helping with the experiments.

References

1. Zhang, L.; Zhou, Y.; Guo, L.; Zhao, W.; Barnes, A.; Zhang, H.-T.; Eaton, C.; Zheng, Y.; Brahlek, M.; Haneef, H. F.; Podraza, N. J.; Chan, Moses H. W.; Gopalan, V.; Rabe, K. M.; Engel-Herbert, R., Correlated metals as transparent conductors. *Nature Materials* **2016,** *15* (2), 204-210.
2. Stoner, J. L.; Murgatroyd, P. A. E.; O'Sullivan, M.; Dyer, M. S.; Manning, T. D.; Claridge, J. B.; Rosseinsky, M. J.; Alaria, J., Chemical Control of Correlated Metals as Transparent Conductors. *Advanced Functional Materials* **2019,** *29* (11), 1808609.
3. Ha, Y.; Byun, J.; Lee, J.; Lee, S., Design Principles for the Enhanced Transparency Range of Correlated Transparent Conductors. *Laser & Photonics Reviews* **2021,** *15* (5), 2000444.
4. Ofoegbuna, T.; Peterson, B.; da Silva Moura, N.; Nepal, R.; Kizilkaya, O.; Smith, C.; Jin, R.; Plaisance, C.; Flake, J. C.; Dorman, J. A., Modifying Metastable $Sr_{1-x}BO_{3-\delta}$ (B = Nb, Ta, and Mo) Perovskites for Electrode Materials. *ACS Applied Materials & Interfaces* **2021,** *13* (25), 29788-29797.
5. Mohammadi, M.; Xie, R.; Hadaeghi, N.; Radetinac, A.; Arzumanov, A.; Komissinskiy, P.; Zhang, H.; Alff, L., Tailoring Optical Properties in Transparent Highly Conducting Perovskites by Cationic Substitution. *Advanced Materials* **2023,** *35* (7), 2206605.
6. Zhang, X.; Zhang, L.; Perkins, J. D.; Zunger, A., Intrinsic Transparent Conductors without Doping. *Physical Review Letters* **2015,** *115* (17), 176602.
7. Gao, J.; Kempa, K.; Giersig, M.; Akinoglu, E. M.; Han, B.; Li, R., Physics of transparent conductors. *Advances in Physics* **2016,** *65* (6), 553-617.
8. Mizoguchi, H.; Kitamura, N.; Fukumi, K.; Mihara, T.; Nishii, J.; Nakamura, M.; Kikuchi, N.; Hosono, H.; Kawazoe, H., Optical properties of $SrMoO_3$ thin film. *Journal of Applied Physics* **2000,** *87* (9), 4617-4619.
9. Nekrasov, I. A. a. K., G. and Kondakov, D. E. and Kozhevnikov, A. V. and Pruschke, Th. and Held, K. and Vollhardt, D. and Anisimov, V. I., Explanation of the similarity of the experimental photoemission spectra of $SrVO_3$ and $CaVO_3$. *arXiv* **2002**.
10. Mirjolet, M.; Kataja, M.; Hakala, T. K.; Komissinskiy, P.; Alff, L.; Herranz, G.; Fontcuberta, J., Optical Plasmon Excitation in Transparent Conducting $SrNbO_3$ and $SrVO_3$ Thin Films. *Advanced Optical Materials* **2021,** *n/a* (n/a), 2100520.
11. Ok, J. M.; Mohanta, N.; Zhang, J.; Yoon, S.; Okamoto, S.; Choi, E. S.; Zhou, H.; Briggeman, M.; Irvin, P.; Lupini, A. R.; Pai, Y.-Y.; Skoropata, E.; Sohn, C.; Li, H.; Miao, H.; Lawrie, B.; Choi, W. S.; Eres, G.; Levy, J.; Lee, H. N., Correlated oxide Dirac semimetal in the extreme quantum limit. *Science Advances* **2021,** *7* (38), eabf9631.
12. Bigi, C.; Orgiani, P.; Sławińska, J.; Fujii, J.; Irvine, J. T.; Picozzi, S.; Panaccione, G.; Vobornik, I.; Rossi, G.; Payne, D.; Borgatti, F., Direct insight into the band structure of $SrNbO_3$. *Physical Review Materials* **2020,** *4* (2), 025006.

13. Roth, J.; Paul, A.; Goldner, N.; Pogrebnyakov, A.; Agueda, K.; Birol, T.; Alem, N.; Engel-Herbert, R., Sputtered SrxNbO3 as a UV-Transparent Conducting Film. *ACS Applied Materials & Interfaces* **2020,** *12* (27), 30520-30529.

14. Park, Y.; Roth, J.; Oka, D.; Hirose, Y.; Hasegawa, T.; Paul, A.; Pogrebnyakov, A.; Gopalan, V.; Birol, T.; Engel-Herbert, R., SrNbO$_3$ as a transparent conductor in the visible and ultraviolet spectra. *Communications Physics* **2020,** *3* (1), 102.

15. Oka, D.; Hirose, Y.; Nakao, S.; Fukumura, T.; Hasegawa, T., Intrinsic high electrical conductivity of stoichiometric SrNbO$_3$ epitaxial thin films. *Physical Review B* **2015,** *92* (20), 205102.

16. Radetinac, A.; Takahashi, K. S.; Alff, L.; Kawasaki, M.; Tokura, Y., Single-Crystalline CaMoO$_3$ and SrMoO$_3$ Films Grown by Pulsed Laser Deposition in a Reductive Atmosphere. *Applied Physics Express* **2010,** *3* (7), 073003.

17. Wan, D.; Yan, B.; Chen, J.; Wu, S.; Hong, J.; Song, D.; Zhao, X.; Chi, X.; Zeng, S.; Huang, Z.; Li, C.; Han, K.; Zhou, W.; Cao, Y.; Rusydi, A.; Pennycook, S. J.; Yang, P.; Ariando; Xu, R.; Xu, Q.-H.; Wang, X. R.; Venkatesan, T., New Family of Plasmonic Photocatalysts without Noble Metals. *Chemistry of Materials* **2019,** *31* (7), 2320-2327.

18. Yang, L.; Yu, J.; Fu, Q.; Kong, L.; Xu, X., Mesoporous single-crystalline SrNbO2N: Expediting charge transportation to advance solar water splitting. *Nano Energy* **2022,** *95*, 107059.

19. Wan, D. Y.; Zhao, Y. L.; Cai, Y.; Asmara, T. C.; Huang, Z.; Chen, J. Q.; Hong, J.; Yin, S. M.; Nelson, C. T.; Motapothula, M. R.; Yan, B. X.; Xiang, D.; Chi, X.; Zheng, H.; Chen, W.; Xu, R.; Ariando; Rusydi, A.; Minor, A. M.; Breese, M. B. H.; Sherburne, M.; Asta, M.; Xu, Q. H.; Venkatesan, T., Electron transport and visible light absorption in a plasmonic photocatalyst based on strontium niobate. *Nature Communications* **2017,** *8* (1), 15070.

20. Asmara, T. C.; Wan, D.; Zhao, Y.; Majidi, M. A.; Nelson, C. T.; Scott, M. C.; Cai, Y.; Yan, B.; Schmidt, D.; Yang, M.; Zhu, T.; Trevisanutto, P. E.; Motapothula, M. R.; Feng, Y. P.; Breese, M. B. H.; Sherburne, M.; Asta, M.; Minor, A.; Venkatesan, T.; Rusydi, A., Tunable and low-loss correlated plasmons in Mott-like insulating oxides. *Nature Communications* **2017,** *8* (1), 15271.

21. Garcia-Castro, A. C.; Ma, Y.; Romestan, Z.; Bousquet, E.; Cen, C.; Romero, A. H., Engineering of Ferroic Orders in Thin Films by Anionic Substitution. *Advanced Functional Materials* **2022,** *32* (2), 2107135.

22. Alff, L.; Klein, A.; Komissinskiy, P.; Kurian, J., Vapor-Phase Deposition of Oxides. In *Ceramics Science and Technology*, Riedel, I.-W. C. a. R., Ed. Wiley-VCH Verlag GmbH: 2012; Vol. 3, pp 267-290.

23. Palakkal, J.; Meyer, T.; Major, M.; Bierschenk, L. E.; Alff, L., Growth Engineering of SrNbO$_3$ Perovskite Oxide by Pulsed Laser Deposition and Molecular Beam Epitaxy. *arXiv* **2024**, 17.

24. Palakkal, J. P.; Schneider, T.; Alff, L., Oxygen defect engineered magnetism of La$_2$NiMnO$_6$ thin films. *AIP Advances* **2022,** *12* (3), 035116.

25. Palakkal, J.; Henning, P.; Alff, L., Effect of valence electrons on the core level x-ray photoelectron spectra of niobium oxide thin films prepared by molecular beam epitaxy. *arXiv* **2024**, 18.

26. Johs, B.; Hale, J. S., Dielectric function representation by B-splines. *physica status solidi (a)* **2008,** *205* (4), 715-719.

27. van Benthem, K.; Elsässer, C.; French, R. H., Bulk electronic structure of SrTiO3: Experiment and theory. *Journal of Applied Physics* **2001,** *90* (12), 6156-6164.

28. Madsen, G. K. H.; Blaha, P.; Schwarz, K.; Sjöstedt, E.; Nordström, L., Efficient linearization of the augmented plane-wave method. *Physical Review B* **2001,** *64* (19), 195134.

29.     Schwarz, K.;  Blaha, P.; Madsen, G. K. H., Electronic structure calculations of solids using the WIEN2k package for material sciences. *Computer Physics Communications* **2002,** *147* (1), 71-76.
30.     Blaha, P.;  Schwarz, K.;  Tran, F.;  Laskowski, R.;  Madsen, G. K. H.; Marks, L. D., WIEN2k: An APW+lo program for calculating the properties of solids. *The Journal of Chemical Physics* **2020,** *152* (7), 074101.
31.     Perdew, J. P.;  Burke, K.; Ernzerhof, M., Generalized Gradient Approximation Made Simple. *Physical Review Letters* **1996,** *77* (18), 3865-3868.
32.     Koepernik, K.; Eschrig, H., Full-potential nonorthogonal local-orbital minimum-basis band-structure scheme. *Physical Review B* **1999,** *59* (3), 1743-1757.